\documentclass[final]{raa}            

\usepackage{graphicx,times}             
\usepackage{natbib}
\usepackage{amssymb,amsmath}
\bibpunct{(}{)}{;}{a}{}{,}

\newcommand{\cm}{{~\rm cm}}

\newcommand{\km}{{~\rm km}}
\newcommand{\s}{{~\rm s}}

\newcommand{\yr}{{~\rm yr}}

\newcommand{\Gyr}{{~\rm Gyr}}
\newcommand{\pc}{{~\rm pc}}

\begin{document}

   \title{Common envelope to explosion delay time distribution (CEEDTD) of type Ia supernovae
}

   \volnopage{Vol.0 (20xx) No.0, 000--000}      
   \setcounter{page}{1}          

   \author{Noam Soker
      \inst{1}
   }

   \institute{Department of Physics, Technion, Haifa, 3200003, Israel;  soker@physics.technion.ac.il {\it soker@physics.technion.ac.il}\\
\vs\no
   {\small Received~~20xx month day; accepted~~20xx~~month day}}

\abstract{
I use recent observations of circumstellar matter (CSM) around type Ia supernovae (SNe Ia) to estimate the fraction of SNe Ia that explode into a planetary nebula (PN) and to suggest a new delay time distribution from the common envelope evolution (CEE) to the SN Ia explosion for SNe Ia that occur shortly after the CEE.  Under the assumption that the CSM results from a CEE, I crudely estimate that about 50 per cent of all SNe Ia are SNe Ia inside PNe (SNIPs), and that the explosions of most SNIPs occur within a CEE to explosion delay (CEED) time of less than about ten thousand years. I also estimate that the explosion rate of SNIPs, i.e., the CEED time distribution (CEEDTD), is roughly constant within this time scale of ten thousand years.  The short CEED time suggests than a fraction of SNIPs come from the core-degenerate (CD) scenario where the merger of the core with the white dwarf takes place at the end of the CEE. I present my view that the majority of SNIPs come from the CD scenario.  I list some further observations that might support or reject my claims, and the challenge to theoretical studies to find a process to explain a merger to explosion delay (MED) time of up to ten thousand years or so.  A long MED will apply also to the double degenerate scenario. 
}

 \authorrunning{N. Soker}            
   \titlerunning{Common envelope to explosion delay time of SNe Ia}  
   
      \maketitle
\section{INTRODUCTION}
\label{sec:intro}

There is no consensus neither on the main scenario that brings white dwarfs (WDs) to ignite thermonuclear explosions as type Ia supernovae (SNe Ia), nor on the classification of the different SN Ia scenarios 
(e.g., \citealt{LivioMazzali2018, Wang2018,  Jhaetal2019NatAs, RuizLapuente2019, Soker2019Rev, Ruiter2020} for very recent reviews). 
As most recent studies continue to explore all scenarios (e.g. \citealt{Pan2020, Wuetal2020,  Blondinetal2021, Clarketal2021, Ferrandetal2021, LivnehKatz2020, Ablimit2021, LivnehKatz2021, Liuetal2021, MengLuo2021, Michaely2021, Zengetal2021, Chandraetal2021, Patraetal2021}), I briefly review (by alphabetical order) the six binary SN Ia scenarios according to \cite{Soker2019Rev}, where one can find more details. I will not discuss single star scenarios (e.g., \citealt{Clavelli2019, Antoniadisetal2020}).  

Before I list the scenarios I define the merger to explosion delay (MED) time that I introduced in \cite{Soker2018Rev}. The MED time is the time from the merger (or mass transfer) event to the explosion itself (more in section \ref{subsec:MED}). 

(1) The \textit{core-degenerate (CD) scenario}. In this scenario the merger process of a CO WD (or possibly HeCO WD)  with the CO core (or possibly HeCO core) of a massive asymptotic giant branch (AGB) star takes place at the end of the common envelope evolution (CEE) and forms a WD remnant with a mass close to the Chandrasekhar mass limit (e.g., \citealt{KashiSoker2011, Ilkov2013, AznarSiguanetal2015}).  In this scenario a MED time is built-in, and exists in all SNe Ia by this scenario. Its value can be $ 0 < t_{\rm MED} \la 10^{10} \yr$. However, as I argue in the present study (section \ref{subsec:EstimatingCEEDTD}) there is a high SN Ia explosion rate for  $t_{\rm MED} \la 10^{4} \yr$. 
The largest challenges of the CD scenario are to show that core-WD mergers can lead to a large population of WD remnants with masses close to the Chandrasekhar mass limit, and that these remnants have long MED times. The new study by \cite{Neopaneetal2022} of WD-WD mergers suggests that these requirements are possible to reach. 

(2+3) The \textit{double degenerate (DD) scenarios.} In these scenarios (e.g., \citealt{Webbink1984, Iben1984}) the two WDs merge as they lose energy to gravitational waves. One or two of the WDs might be HeCO WDs rather than pure CO WDs
(e.g., \citealt{YungelsonKuranov2017, Zenatietal2019, Peretsetal2019}).
In the \textit{DD scenario} without MED the explosion is likely to take place during a violent merger process (e.g., \citealt{Pakmoretal2011, Liuetal2016, Ablimitetal2016}). In the \textit{DD-MED scenario} the explosion occurs a long time, more than several months, after the merger process. The value of the MED time is an open question of the DD scenarios (e.g., \citealt{LorenAguilar2009, vanKerkwijk2010, Pakmor2013, Levanonetal2015, LevanonSoker2019}).

(4) The \textit{double-detonation (DDet) scenario.} In this scenario the companion transfers helium-rich gas to a CO or a HeCO WD. The accumulated helium layer ignites and explodes, sending a shock wave into the mass-accreting WD and explodes it (e.g., \citealt{Woosley1994, Livne1995, Shenetal2018}). This scenario has no MED time. 

(5) The \textit{single degenerate (SD) scenario.} In this scenario the WD accretes a hydrogen-rich material from a non-degenerate companion and explodes as a (close to) Chandrasekhar-mass WD  (e.g., \citealt{Whelan1973, Han2004, Wangetal2009}). It might have no delay time, i.e., explodes as soon as it reaches the Chandrasekhar mass (or very close to it) or much later after it loses angular momentum (e.g., \citealt{Piersantietal2003, DiStefanoetal2011, Justham2011}),  i.e., it has a MED. In the CEE-wind SD scenario \citep{MengPodsiadlowski2017} the explosion might take place shortly after a CEE in case the WD is a hybrid CONe WD \citep{MengPodsiadlowski2018}. In my recent review  
\cite{Soker2019Rev} I argued that SD-MED, i.e., the SD scenario with MED time, might account for a small fraction of SNe Ia, while in cases of an explosion at the moment the WD reaches close the the Chandrasekhar mass the outcome is a peculiar SN Ia. 

(6) The \textit{WD-WD collision (WWC) scenario.} The collision of two WDs with each other at about their free fall velocity sets an explosion (e.g., \citealt{Raskinetal2009, Rosswogetal2009, Kushniretal2013, AznarSiguanetal2014}). 
Studies argue that this scenario might at best supply $<1 \%$ of all SNe Ia (e.g., \citealt{Toonenetal2018, HallakounMaoz2019, HamersThompson2019}). This scenario has no MED time. 

Some of the scenarios involve a CEE phase. In those scenarios in addition to the MED time there is the CEE to explosion delay (CEED) time. In \cite{Soker2019CEEDTD} I introduced the CEE to explosion delay time distribution (CEEDTD), and wrote an approximate expression for it. In the present study I use new observations from the literature (section \ref{sec:Short}) to propose a new  expression for the CEEDTD (section \ref{subsec:EstimatingCEEDTD}). In deriving this expression I assume that the CSM results from a CEE rather than from a wind from a giant star or from a mass transfer episode.  In section \ref{sec:summary} I summarise the expression for the CEEDTD in the frame of the CD scenario,  

\section{Definitions of the delay times}
\label{sec:DelayTimes}

\subsection{The delay time distribution (DTD)}
\label{subsec:delay}

The delay time distribution (DTD) refers to the distribution of the delay time from star formation to the SN Ia explosion
\begin{equation}
t_{\rm SF-E}\equiv {\rm Star~formation~to
~explosion}.
\label{eq:tSFE}
\end{equation}
Different groups obtain somewhat different DTDs (e.g., \citealt{Grauretal2014, Heringeretal2017, MaozGraur2017, Frohmaieretal2019, Wisemanetal2021}). I use here the same expression as I derived in \cite{Soker2019CEEDTD} that I based on the DTDs of \cite{FriedmannMaoz2018} for galaxy clusters and of \cite{Heringeretal2019} for field galaxies 
\begin{equation}
\dot N_{\rm DTD}  = 0.19 N_{\rm Ia} F_1(t_{\rm i}) \left( \frac{t_{\rm SF-E}}{1 \Gyr} \right)^{-1.32} \Gyr^{-1},
\label{eq:dotN3A}
\end{equation}
where 
\begin{equation}
F_1(t_{\rm i})=1.68 \left[ (t_{\rm i}/\Gyr)^{-0.32} - 13.7^{-0.32} \right]^{-1},
\label{eq:dotN3B}
\end{equation}
and SNe Ia occur in the time interval $t_{\rm i} < t_{\rm SF-E} < 13.7 \Gyr$. Namely, $t_{\rm i}$ is the first time after star formation when SNe Ia occur. 

\cite{FriedmannMaoz2018} consider $t_{\rm i} = 0.04 \Gyr$ and find a SN Ia efficiency, i.e., number of SNe Ia per formed stellar mass to be $n_{\rm Ia}\simeq 0.003-0.008 M^{-1}_\odot$. 
\cite{Heringeretal2019} have $t_{\rm i} = 0.1 \Gyr$ and find $n_{\rm Ia} \simeq 0.003-0.006 M^{-1}_\odot$. To proceed I take here $t_{\rm i}=0.05 \Gyr$. This value  corresponds to the life time of star of initial mass of $M_{\rm ZAMS} \simeq 6 M_\odot$ from the main sequence to its asymptotic giant branch phase. Substituting $t_{\rm i}=0.05 \Gyr$ in equations (\ref{eq:dotN3A}) and (\ref{eq:dotN3B}) yields 
\begin{equation}
\begin{aligned}
\dot N_{\rm DTD} &  =  0.147 N_{\rm Ia} \left( \frac{t_{\rm SF-E}}{1 \Gyr} \right)^{-1.32} \Gyr^{-1} 
\\ &
 {\rm for} \quad 0.05 \Gyr < t_{\rm SF-E} < 13.7 \Gyr. 
\label{eq:dotNf}
\end{aligned}
\end{equation}
This expression has large uncertainties, but it is not the focus of this study. 

I emphasise that the DTD (\ref{eq:dotNf}) includes \textit{all SNe Ia}, including those with very short delay time after CEE that I discuss below. 

\subsection{Merger to explosion delay (MED) time}
\label{subsec:MED}

This section is relevant mainly to the CD and DD scenarios that have merger of two degenerate stars. For the present study I define the MED time as
\begin{equation}
t_{\rm MED}\equiv {\rm Merger~to
~explosion}.
\label{eq:tMED}
\end{equation}
This time refers also the time from the termination of the mass transfer to explosion in the SD scenario \citep{Soker2018Rev},  which, however, is not the focus of this study. 

In earlier papers (\citealt{Soker2018Rev, Soker2019Rev, Soker2019CEEDTD}) I presented arguments for  that many SNe Ia must have a substantial time delay from the merger event to the explosion. 
The main one is that prompt explosions in the DD scenario lead to highly-non-spherical explosions 
(e.g.,  \citealt{Kashyapetal2017, Pakmor2012, Tanikawaetal2015, vanRossumetal2016}), which is in contradiction with the morphlogy of most SN Ia remnants (SNRs Ia) that tend to be spherical or axisymmetrical (e.g., \citealt{Lopezetal2011}). 
A substantial MED time, i.e., one that is much longer than the dynamical time of the merger process and therefore allows the formation of a single WD, overcomes this challenge. 
The DDet scenario also leads to non-spherical explosions (e.g., \citealt{Papishetal2015, Tanikawaetal2018, Tanikawaetal2019}), but there is no possibility in the DDet scenario to overcome this challenge as it has no MED time.   

\subsection{Common envelope to explosion delay (CEED) time}
\label{subsec:CEEDtime}

In \cite{Soker2019CEEDTD} I listed my motivations to define the CEE to explosion delay (CEED) time 
\begin{equation}
t_{\rm CEED} \equiv {\rm End~of~CEE~to~explosion}. 
\label{eq:tCEED}
\end{equation}
The first motivation comes from the interaction of the Kepler SNR Ia with a massive circumstellar matter (CSM; e.g., \citealt{Sankritetal2016}). The CSM comes from a CEE as there are no indications for any giant star that could have blown the CSM (e.g., \citealt{Kerzendorfetal2014, Medanetal2017}). This suggests that the CSM was blown during a CEE in the frame of the CD scenario or of the DD scenario \citep{Soker2019CEEDTD}.  \cite{MengLi2019} propose the CEE-wind channel of the SD scenario for the Kepler SNR, but I will not discuss it here. 
Another motivation is my view \citep{Sokeretal2013} that the CSM of the SN PTF11kx is too massive for the SD scenario to account for, beside the CEE-wind channel of the SD scenario \citep{MengPodsiadlowski2018}.

In the present study (section \ref{subsec:fraction}) I consider new observations of more SNR Ia that have massive CSM and use these to better constrain the CEED time distribution (CEEDTS). 

As the duration of the CEE might be long relative to $t_{\rm MED}$ and $t_{\rm CEED}$ it is important to refer to the starting time of $t_{\rm CEED}$.  I take $t_{\rm CEED}=0$ at the end of the CEE. In the CD scenario this is also the merger time, i.e., in the CD scenario 
\begin{equation}
t_{\rm CEED} = t_{\rm MED} \qquad {\rm for~the~CD~scenario}. 
\label{eq:CEEDMEDcd}
\end{equation}
A comment on equality of equation (\ref{eq:CEEDMEDcd}) is in place here. In principle the merger can take place before the system ejects the common envelope. However, the merger itself releases a large amount of energy that is likely to eject the entire leftover envelope. In principle the merger might take place after the last amount of bound envelope mass forms a circumbinary disk around the WD-core close binary system \citep{KashiSoker2011}. I here consider the phase of the circumbinary disk to be part of the CEE. Overall, the most likely case is that the core-WD merger takes place at the same time as the termination of the CEE. 

In the DD scenario the time from the end of the CEE to explosion includes the time to merger due to gravitational waves plus the MED time
\begin{equation}
t_{\rm CEED} = t_{\rm GW} + t_{\rm MED} \qquad {\rm for~the~DD~scenario}. 
\label{eq:CEEDMEDdd}
\end{equation}

\section{The SNe Ia rate shortly after the CEE}
\label{sec:Short}

\subsection{The fraction of SNe Ia inside PNe (SNIPs)}
\label{subsec:fraction}

\subsubsection{Previous estimates}
\label{subsubsubsec:previous}

\cite{TsebrenkoSoker2015a} assumed that the presence of two opposite protrusions from the main shell of a SNR Ia, termed `Ears', indicates that the ejecta interacts with a CSM that once was a planetary nebula (PN)  (see also \citealt{Chiotellisetal2016}). \cite{TsebrenkoSoker2015a} assumed that the ears are along the symmetry axis, i.e., polar ears, and named these SNe Ia SNIPs, for SNe inside PNe, including also SNe Ia explosions inside proto-PNe \citep{Cikotaetal2017}. I note that \cite{Chiotellisetal2021} study a model where the ears are in the equatorial plane rather than at the polar directions of the SNR. 
\cite{TsebrenkoSoker2015a} examined the morphology of 13 SNRs and from that estimated that SNIPs amount to at least $\simeq 20 \pm 10 \%$ of all SNe Ia. They derived this number from their estimate, by the presence of ears, that two SNRs Ia posses ears while four other SNRs Ia might posses ears. 
I list these SNRs in Table \ref{table:SNRs}, indicating the classification of \cite{TsebrenkoSoker2015a} in the third row. 
\begin{table*}[t]
        \centering
    \begin{tabular}{|lccc|cccc|}
    \hline
    SNR & X-ray Image & `Ears' & This study & SNR  & X-ray Image & `Ears' & This study       \\
    \hline
    Kepler     &  \includegraphics[width=15mm, height=15mm]{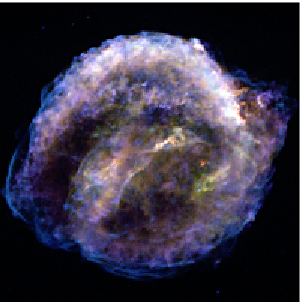}     & Yes & Yes (Ears + CSM)    
    & SN~1006         &  \includegraphics[width=15mm, height=15mm]{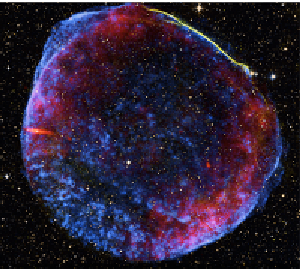}       & No & Yes (CSM) \\
    G1.9+0.3   &  \includegraphics[width=15mm, height=15mm]{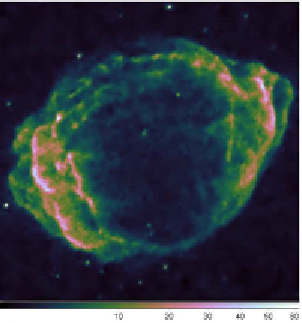}   & Yes &  Yes (Ears) 
    & 3C~397   &  \includegraphics[width=15mm, height=15mm]{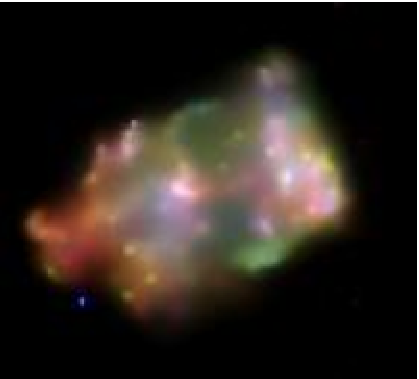} & No  & Maybe \\
    G299.2-2.9 &  \includegraphics[width=15mm, height=15mm]{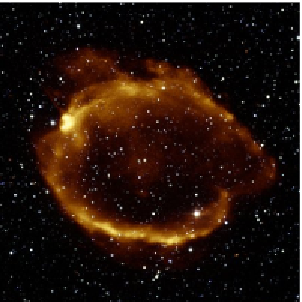}       & Maybe  & Yes (Ears)   
    & Tycho &  \includegraphics[width=15mm, height=15mm]{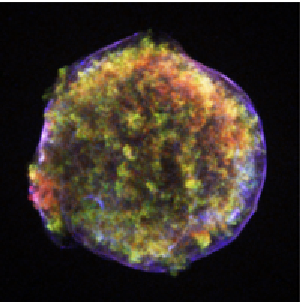}     & No & Maybe (CSM) \\
    RCW86      &  \includegraphics[width=15mm, height=15mm]{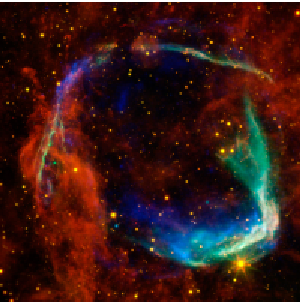}      & No  & Yes (Ears + CSM)  
    &   DEM~L71      &  \includegraphics[width=15mm, height=15mm]{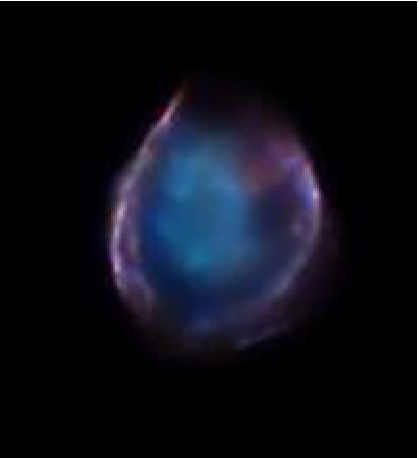}   & Maybe  & Yes (Ears + CSM) \\
    N~103B      &  \includegraphics[width=15mm, height=15mm]{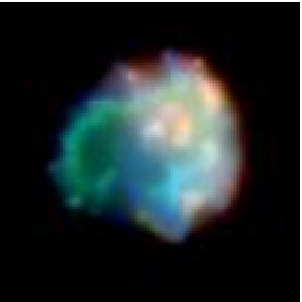}        & No  & Yes (CSM)    
    & 0548-70.4 &  \includegraphics[width=15mm, height=15mm]{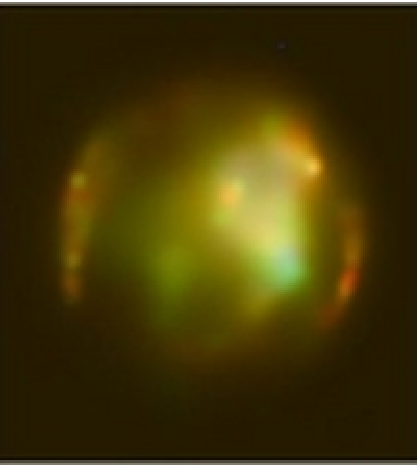}      & No  & Yes (Ears + CSM) \\
    0534$-$69.9  &  \includegraphics[width=15mm, height=15mm]{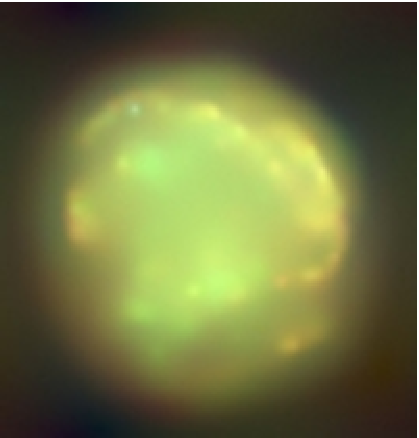}  & Maybe & Maybe (Ears)   
    & 0509-67.5    &  \includegraphics[width=15mm, height=15mm]{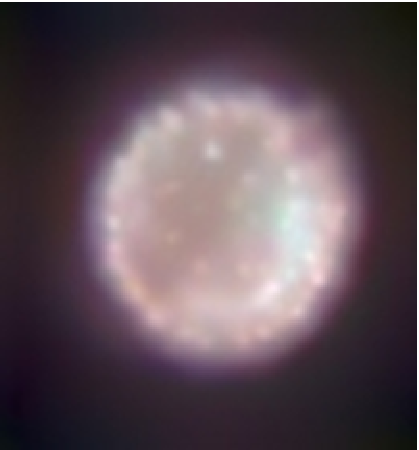}  & No & No \\
    0519-69.0  &  \includegraphics[width=15mm, height=15mm]{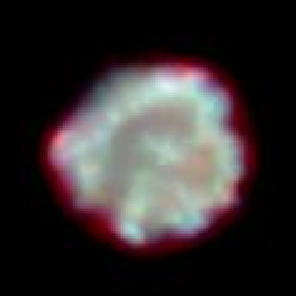}  & Maybe  & Yes (CSM)   
    & ---    &  ---  & --- & ---  \\
    DEM~L249  &  \includegraphics[width=15mm, height=15mm]{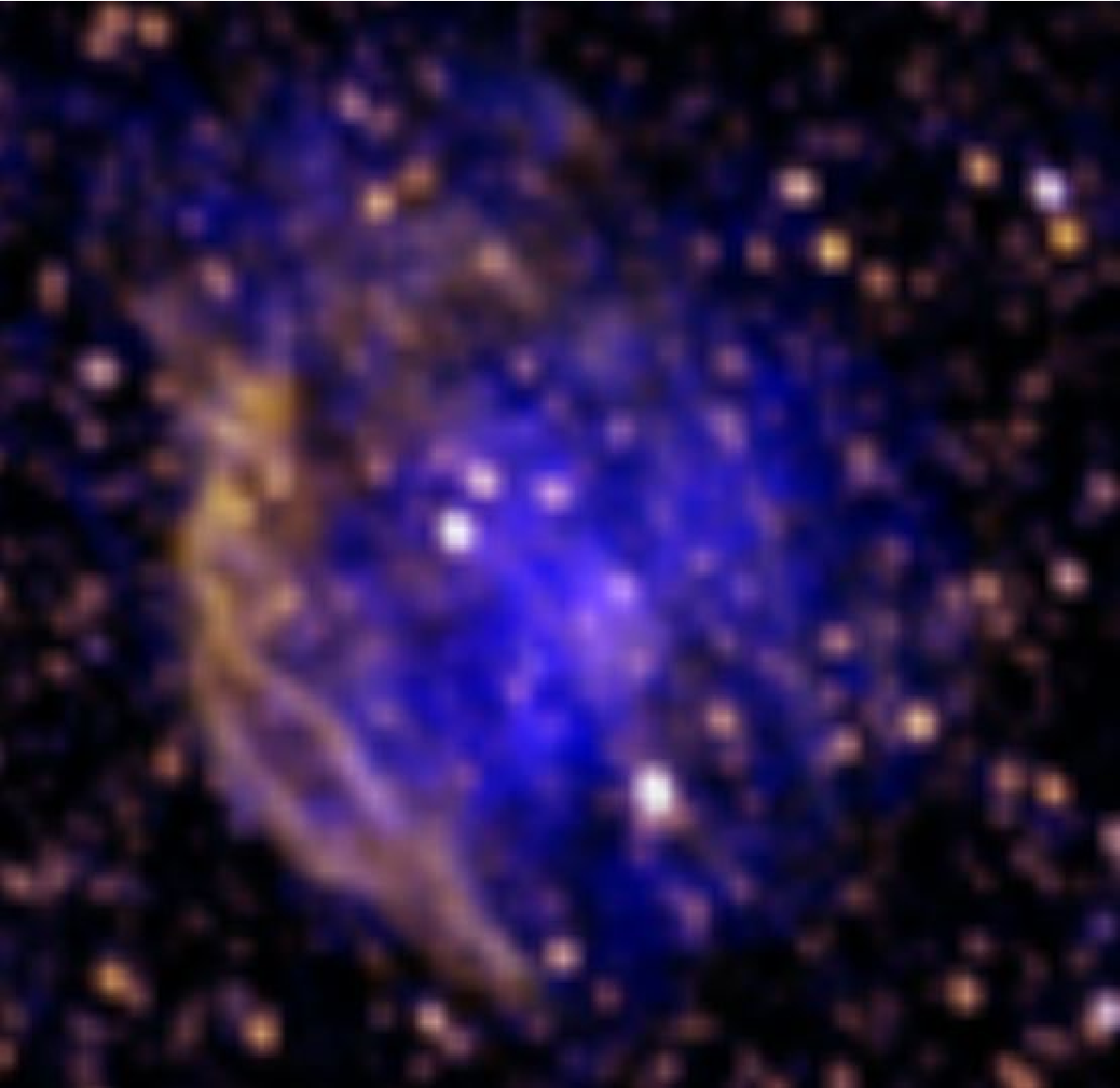}  & ---  & Yes (CSM)     
    & DEM~L238  &  \includegraphics[width=15mm, height=15mm]{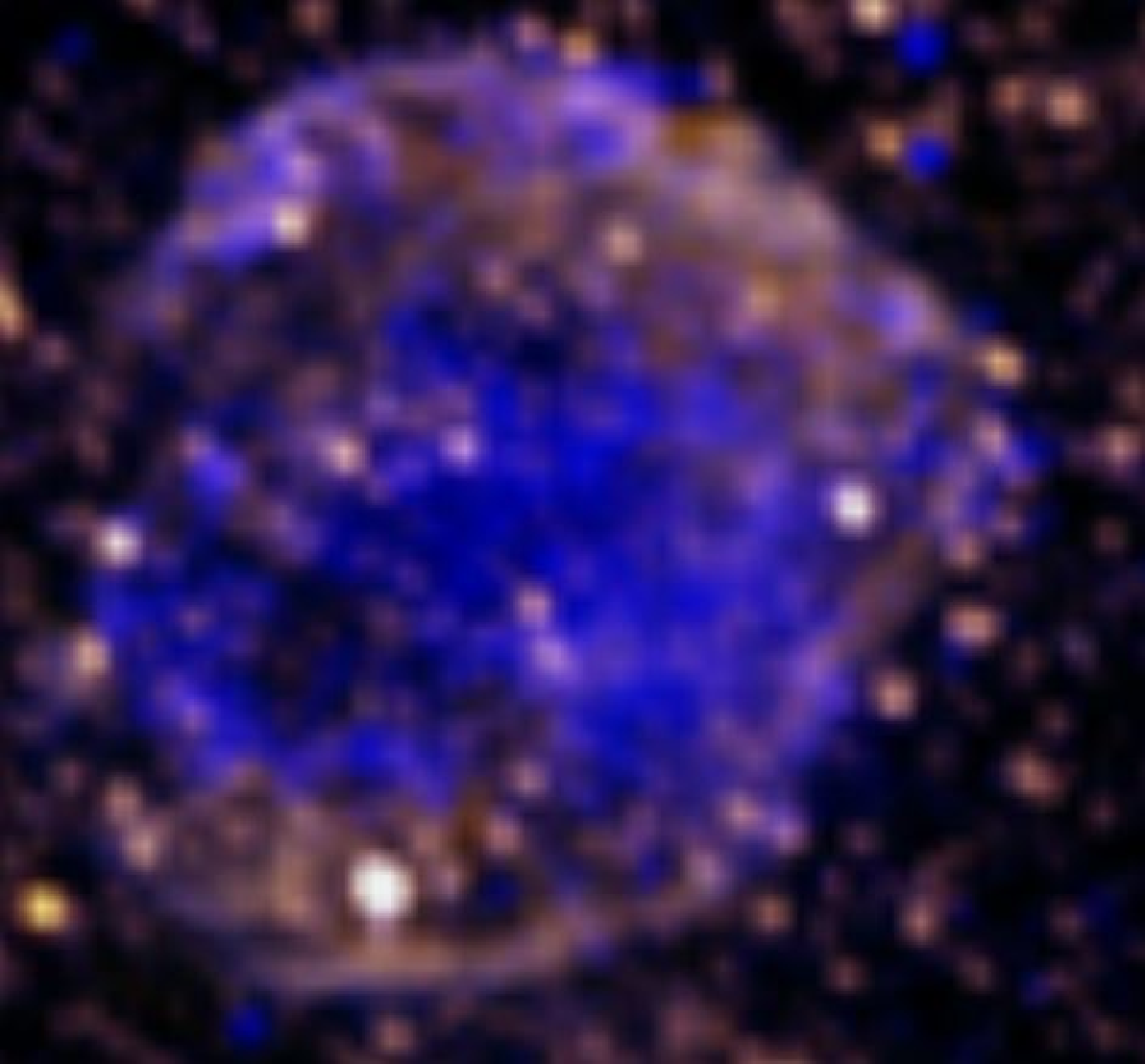}  & ---  & Yes (CSM)    \\
    \hline
    \end{tabular}
    \caption{Known well-resolved SNRs with ages less than $10,000 \yr$ (based on \citealt{TsebrenkoSoker2015a}, beside the two SNe Ia in the bottom row). The third column is the classification by \cite{TsebrenkoSoker2015a} of whether the SNR is a SNIP (a type Ia supernova inside a planetary nebula) or not, according to the presence or not of `ears', i.e., two opposite protrusions. The fourth column lists my new estimate in the present study of whether the SNR is a SNIP or not. The images are from the Chandra SNR Catalogue (references therein). }
      \label{table:SNRs}
\end{table*}

I add two SNe Ia to the list of \cite{TsebrenkoSoker2015a}, DEM~L249 and DEM~L238 (bottom row of Fig. \ref{table:SNRs}) that I take from \cite{Borkowskietal2006}. 
\cite{Borkowskietal2006} suggest the presence of  substantial amounts of dense circumstellar gas at the explosions of these two SNe Ia, and that these SNe Ia could be remnants of prompt SNe Ia, i.e., within $\simeq 10^8 \yr$ from star formation. Based on this estimated dense CSM and the morphologies, I consider these SNe Ia to be SNIPs.

In \cite{Soker2019CEEDTD} I noted that the SNR Ia N103B does interact with a CSM  (e.g., \citealt{Williamsetal2018}). I still estimated the fraction of SNIP as $f_{\rm SNIP} \simeq 15 - 20 \%$. I further assumed that the time delay to SN Ia explosion during which we can detect an interaction with a CSM is $t < t_{\rm SNIP} \simeq 3 \times 10^5 \yr$. From that I derive the average SN Ia rate in the time period $0< t_{\rm CEED} < 3 \times 10^5$ to be $\approx (100-1000)N_{\rm Ia} \Gyr^{-1}$. 
Below I use new observational results to estimate a much higher rate, due both to a larger fraction of SNIPs and to a shorter relevant ejecta-CSM interaction time.  

\subsubsection{New estimates of SNIP fraction}
\label{subsubsubsec:NewSNIP}

\cite{Lietal2021} present a study of the SNe Ia with Balmer-dominated shells in the Large Magellanic Cloud (LMC). One of them is N103B that has a CSM mass of about $1-3 M_\odot$ (e.g., \citealt{Williamsetal2014, Lietal2017, Blairetal2020}). \cite{Lietal2021} find that the LMC SNRs Ia N103B, DEML71, and 0548$-$70.4 have numerous and wide-spread knots, and that their density is too high to be of interstellar medium (ISM) source. These high-density knots most-likely originate from a CSM gas. The LMC SNR Ia 0519$-$69.0 has a small number of knots.
There are other indications for the presence of CSM around some SNe Ia, e.g., by the presence of sodium absorption lines (e.g., \citealt{Patatetal2007Sci, Sternbergetal2011Sci}). For these cases as well, its seems that the CSM is too massive for the SD scenario and better fits the expectations of the CD scenario (e.g., \citealt{Soker2015Na}).  
   
Based on the new study of \cite{Lietal2021} I consider the possibility that the other SNRs Ia that have Balmer-dominated spectra also have a massive CSM, namely, be SNIPs. These are Tycho \citep{KirshnerChevalier1978}, SN~1006 \citep{SchweizerLasker1978}, Kepler \citep{Fesenetal1989}, and RCW86 \citep{LongBlair1990}. As Tycho has a global spherically symmetric morphology, I mark it as maybe for being a SNIP. Of course, not all SNe Ia have CSM  (e.g., \citealt{Cendesetal2020}). With the two addition to the table (bottom row of Table \ref{table:SNRs}), 11-14 out of 15 SNRs Ia are SNIPs.  

From my assessment that 11 out of 15 SNRs Ia are SNIPs and that 3 are `maybe', my new estimate is that in the Galaxy and in the LMC the SNIP fraction is $11/15$ to $(11+3/2)/15$, or 
\begin{equation}
f_{\rm SNIP}{\rm (MW+LMC)} \simeq 0.7-0.8. 
\label{eq:fSNIPMW}
\end{equation}
However, Table \ref{table:SNRs} overestimates the fraction of SNIPs for several reasons. 
(1) SNIPs have a large amount of CSM, which implies that the interaction of the SN Ia ejecta with the CSM makes the SNR brighter, in X-ray, radio, and optical, hence easier to detect. Namely, we might miss some non-SNIP SNRs Ia. (2) The CSM slows down the expansion of the ejecta. This implies that SNIPs can be detected for a longer time. 
(3) The Galaxy and in the LMC have on-going star formation so I expect a large fraction of SNe Ia with short time delay from star formation, which in the present study implies a large fraction of SNe Ia with short CEED time, i.e., SNIPs. The fraction of SNIPs when we include elliptical galaxies with lower star formation rates is lower. If SNIPs come from near-Chandrasekhar-mass SNe Ia as I claim here, the higher fraction of SNIPs in star-forming galaxies might be compatible with the finding of \cite{Kobayashietal2020} that the fraction of near-Chandrasekhar-mass SNe Ia in the Milky Way is higher than in dwarf galaxies.
\cite{Brownetal2019} find that the specific SN Ia rate is much larger in low-mass galaxies than in massive galaxies. 
This might also point to a very large fraction of SNe Ia shortly after star formation as the low-mass galaxies have a higher specific star formation rate. \cite{Smithetal2012}, for example, argued that the SN Ia rate per unit stellar mass is a positive function of specific star formation rate. 

Because of the large uncertainties  in the degree by which the three effects cause Table \ref{table:SNRs}, as expressed in equation (\ref{eq:fSNIPMW}), to overestimate the SNIP fraction, I simply take 
\begin{equation}
f_{\rm SNIP}{\rm (total)} \simeq 0.5. 
\label{eq:fSNIPtotal}
\end{equation}

If all SNe Ia that take place shortly (see section \ref{subsec:CEEDtime}) after star formation are SNIPs, then from the DTD in equation (\ref{eq:dotNf}) I find that SNe Ia are SNIPs in the time range of 
$0.05 \Gyr < t _{\rm SF-E} < 0.27 \Gyr$. This time span approximately corresponds to star of zero age main sequence masses in the range of 
\begin{equation}
3.5 M_\odot \la M_{\rm ZAMS} \la 6 M_\odot 
\qquad{\rm SNIP ~progenitors} .
\label{eq:MSNIP}
\end{equation}
This range of not-too-low-mass progenitors ensures that in many cases a WD companion merges with the core of an AGB star during the CEE (e.g., \citealt{Sokeretal2013, Ablimitetal2021}). Not in all cases a WD-core merger takes place at the right time. In some cases mass transfer can be stable, or it might take place at very early evolutionary phases, like during the Hertzsprung-Gap (e.g. \cite{Hachisuoetal2008, MengPodsiadlowski2017}). I assume that sufficiently large number of systems to account for SNIPs do merge at the right time  (e.g., \citealt{Sokeretal2013}). 

\subsection{Estimating the CEED time}
\label{subsec:EstimatingCEED}

The high mass loss rate at the end of the AGB might be long, up to $\approx 10^5 \yr$ (e.g., \citealt{Corradietal2003, Michaelyetal2019, Igoshevetal2020, SantanderGarciaetal2021}). 
I quantify the relevant times from the study by \cite{Corradietal2003} of halos of PNe. In PNe with halos the ages of most halos are in the general range of  $t_{\rm halo} \simeq 2\times 10^4 - 8 \times 10^4 \yr$, while the much more non-spherical inner parts of these PNe that presumably the CEE process formed have ages in the general range of $t_{\rm neb} \simeq  2\times 10^3 - 10^4 \yr$.
\cite{Corradietal2003} calculate the ages of the halos by assuming and expansion velocity of $v_{\rm halo} = 15 \km \s^{-1}$. 
Namely, the radii of the halos are in the general range of 
$r_{\rm halo} \simeq  0.3 - 1.2 \pc$, but one halo extends to $2 \pc$. The young inner parts implies that at the termination of the CEE the outer parts of the halo are at $\simeq 1 \pc$. 

The distances from the center (explosion site) of the dense knots that \cite{Lietal2021} attribute to a CSM in their new study of four SNe Ia are $r_{\rm knots} \simeq 1-10 \pc$. As the AGB progenitors stars that engulf their WD companion are massive (equation \ref{eq:MSNIP}), it is indeed possible that some halos will be very large at the time of explosion. 
The explosion itself can be a short time after the termination of the CEE, as the halo is already large. I take the explosions that form SNIPs to be within the time scale of 
 \begin{equation}
t_{\rm SNIP} \approx 10^4 \yr 
\label{eq:tSNIP}
\end{equation}
after the CEE. This is much shorter than the time I assumed in \cite{Soker2019CEEDTD} which was $3 \times 10^5 \yr$.  

From equations (\ref{eq:fSNIPtotal}) and (\ref{eq:tSNIP}) I find the average explosion rate of SNIPs to crudely be  
\begin{equation}
{\overline {\dot N}}_{\rm SNIP} 
= \frac{f_{\rm SNIP} N_{\rm Ia}}{t_{\rm SNIP}} \approx 5 \times 10^4 N_{\rm Ia} \Gyr^{-1}.
\label{eq:SNIPrate}
\end{equation}
This rate is 50 to 500 larger than the rate I estimated in  \cite{Soker2019CEEDTD}. The new value is motivated to large part by the new results of \cite{Lietal2021}. 

There are other supporting indications. Although the delay time from CEE to explosion might be in the frame of the CD scenario (where $t_{\rm CEED} = t_{\rm MED}$; equation \ref{eq:CEEDMEDcd}), or in the frame of the DD scenario (where equation \ref{eq:CEEDMEDdd} holds) or in the frame of the DDet scenario, due to the short time for gravitational waves to operate I assume that most SNIPs in Table \ref{table:SNRs}  are due to the CD scenario.
As well, there are no definitive evidence for surviving companions stars as expected in the DDet and the SD scenarios (e.g., \citealt{Litkeetal2017, Lietal2019}). The non-detection of companions in the SNIPs that I study rule out the process by which a binary system in the frame of the SD scenario experiences a delayed dynamical instability that leads to a high mass loss rate, a scenario proposed by \cite{HanPodsiadlowski2006}.  

The CD scenario implies that after core-WD merger most merger remnants live for up to $t_{\rm SNIP} \approx 10^4 \yr$ before they explode. The life time of many merger products might be much larger though. The merger remnant is a WD of about the Chandrasekhar mass. This implies the existence of a non-negligible population of WDs with masses close to the the Chandrasekhar mass limit. In \cite{BearSoker2018} we examined catalogues of WDs and concluded that there are sufficient number of massive ($ M_{\rm WD} \ga 1.35 M_\odot$) WDs that might potentially explode as SNe Ia in the frame of the CD scenario. In a very recent study \cite{Caiazzoetal2021} report the discovery of a magnetised WD with a mass of $> 1.35 M_\odot$ and argue, along the earlier claim of \cite{BearSoker2018}, that such objects are not rare. 

\subsection{Estimating the CEED time distribution (CEEDTD)}
\label{subsec:EstimatingCEEDTD}

SNe Ia-CSM are rare (e.g., \citealt{Szalaietal2019, Dubayetal2021}) class of SNe Ia that interact with close CSM, i.e., at $R_{\rm CSM} \la 10^{17} \cm$, such as PTF11kx \citep{Dildayetal2012} and SN~2015cp \citep{Grahametal2019}. 
In \cite{Soker2019CEEDTD} I took a CSM expansion velocity of $10 \km \s^{-1}$, and I assumed that this CSM was formed during the CEE. From these I estimates the CEED time of SNe Ia-CSM to be  $t_{\rm CSM} \la 3000 \yr$. However, considering the presence of halos in PNe (section \ref{subsec:EstimatingCEED}) this time can be shorter. I take here $t_{\rm CSM} \la 1000 \yr$.
I do note that the uncertainty is large. For example, if the  inner boundary of the ejected common envelope is of a circumbinary equatorial outflow it might be slower, making the time scale longer. I here follow my earlier assumption of an expansion velocity of $10 \km \s^{-1}$.  
I then adopt the estimate of \cite{Grahametal2019} that the fraction of SNe Ia-CSM from all SNe Ia is $f_{\rm CSM} < 0.06$. 
A similar value I estimate from the results of \cite{Dubayetal2021}. For example, for the moderate-luminosity ejecta-CSM interaction  \cite{Dubayetal2021} estimate the fraction of SNe Ia-CSM that have interaction at $0-500$~days [$500-1000$~days] after discovery to be $f_{\rm CSM} \la 0.073$ [$f_{\rm CSM} \la 0.031$].
In a recent study \cite{Sharmaetal2021} find the fraction of SNe-CSM for CSM within a radius of $0.5\times 10^{16} \cm -10^{16} \cm$ to be about 1 event per 300 SNe Ia. Scaling to a radius of $10^{17} \cm$ gives a rate of 10-20 in 300, or 
$\approx 0.03-0.07$. 
From these three studies I take $f_{\rm CSM} \approx 0.05$. This gives that the rate of SNe Ia-CSM is 
\begin{equation}
{\overline {\dot N}}_{\rm CSM} 
= \frac{f_{\rm CSM} N_{\rm Ia}}{t_{\rm CSM}} \approx 5 \times 10^4   N_{\rm Ia} \Gyr^{-1}.
\label{eq:CSMrate}
\end{equation}
Note that the SNe Ia-CSM are part of the SNIPs. Namely, the fraction of $f_{\rm CSM} \approx 0.05$ is part of the fraction given in equation (\ref{eq:fSNIPtotal}) and not in addition to it.  

A comment is in place here about the inner radius of the CSM. After core-WD merger the WD merger remnant is most likely to blow a very fast and tenuous wind as in young (up to several thousands years) PNe. This wind forms a very low density bubble that for hundreds of years accelerates the previously ejected slow wind (e.g., \citealt{VolkKwok1985}). The inner CSM boundary might reach a velocity of $\approx 30-50 \km \s^{-1}$, and in $1000-10,000 \yr$ a radius of $\approx 10^{17} \cm - 1.5 \times 10^{18} \cm= 0.03-0.5 \pc$.  

Because in \cite{Soker2019CEEDTD} my estimates were that ${\overline {\dot N}}_{\rm CSM} \gg {\overline {\dot N}}_{\rm SNIP}$ I found that crudely the rate of SNe Ia shortly after the CEE decreases as $\approx t^{-1}$. However, in this study I find ${\overline {\dot N}}_{\rm CSM} \approx {\overline {\dot N}}_{\rm SNIP}$ (equations \ref{eq:SNIPrate} and \ref{eq:CSMrate}). This brings me to the main result of this study which is a new estimate of the CEE to explosion delay (CEED) time distribution (CEEDTD) shortly after the CEE
\begin{equation}
\begin{aligned}
\dot N_{\rm Ia,short}  \approx & 5 \times 10^4 N_{\rm Ia} \Gyr ^{-1} \\  & {\rm for} \quad t_{\rm CEED} \la 10^4 \yr. 
\label{eq:NewexpFinal}
\end{aligned}
\end{equation}

The following comments are in place here regarding the CEEDTD of equation (\ref{eq:NewexpFinal}). 
\begin{enumerate}
\item The short CEED time ($t_{\rm CEED} \la 10^4 \yr$) means that the expression (\ref{eq:NewexpFinal}) refers to SNe Ia that take place inside PNe, i.e., SNIPs. This includes also SNe Ia-CSM. 
\item The total number of SNe Ia that the short CEEDTD in equation (\ref{eq:NewexpFinal}) refers to is $\approx 0.5 N_{\rm Ia}$. This is part of the total number of SNe Ia as the DTD in equation (\ref{eq:dotNf}) refers to. Namely, the number of SNe Ia that might take place within a time of $t_{\rm CEED} \la 10^4 \yr$ from the CEE is, with large uncertainties, crudely equals to the number of SNe Ia with much larger CEED times.
\item The different dependence on time of the DTD (equation \ref{eq:dotNf}) and of he CEEDTD (equation \ref{eq:NewexpFinal}) is not in contradiction because the time in the DTD is measured from star formation, $t_{\rm SF-E}$, and it is a fit over a very long time of $\simeq 10 \Gyr$, while in the CEEDTD the time is measured from the termination of the CEE.   
\item The short CEEDTD (equation \ref{eq:NewexpFinal}) predicts also explosion within a short time after merger, possibly even before the common envelope is ejected to large distances ($r \ga 10^{14} \cm$),  hence forming a peculiar type II supernova. However, these are very rare.  
\item The short CEEDTD (equation \ref{eq:NewexpFinal}) holds for SN Ia scenarios that require a CEE phase followed by strong interaction or merger, i.e., the CD, DD, and DDet scenario. However, because of the very short time relative to orbital decay due to the emission of gravitational waves I consider the CD scenario as the dominant one for SNe Ia with short CEED times. 
\item If these SNIPs result from the CD scenario it implies that core-WD merger forms WD remnants that are close to the Chandrasekhar mass limit and explode only after a time of up to $\approx 10^4 \yr$. This challenges theoretical studies to come up with a mechanism to delay explosion for such time scales, and possibly to much longer times scales but at a decreasing explosion rate. Most likely this mechanism is applicable also to the DD scenario.  
\end{enumerate}

\section{Summary}
\label{sec:summary}

The new measurements of dense CSM in four SNe Ia in the LMC \citep{Lietal2021} brought me to reassess the fraction of SNe Ia inside PNe, i.e., SNIPs.
Based on the morphological features of two opposite `ears' and the presence of massive CSM (fourth row of Table \ref{table:SNRs}) I estimated this fraction to be $\approx 50 \%$ of all SNe Ia (equation \ref{eq:fSNIPtotal}). This is about twice as large as in an earlier study of the CEEDTD \citep{Soker2019CEEDTD}. By comparing the sizes of the CSM of these SNe Ia with the sizes of halos of some PNe I further estimated that the SN Ia explosions in these and similar SNe Ia took place within a time of $t < t_{\rm SNIP} \simeq 10^4 \yr$ after the end of the CEE. This is about 30 times shorter than my previous estimate in \cite{Soker2019CEEDTD}.
From these I estimated the rate of SNIPs as I give in equation (\ref{eq:SNIPrate}), and found it to be similar to the rate of SNe Ia that interact with close CSM, termed SNe Ia-CSM, as I gave in equation (\ref{eq:CSMrate}). I used these two about equal rates to suggest a new 
CEEDTD in equation (\ref{eq:NewexpFinal}).

Equation (\ref{eq:NewexpFinal}) that describes my new estimate of the delay time distribution of SNe Ia that occur shortly after the termination of the CEE, i.e., the CEEDTD, is the main result of this study. The number of such short-CEEDTD SNe Ia amounts to about half of all SNe Ia. As the time delay is short relative to the time of spiralling-in due to emission of gravitational waves that the DD scenario requires, I claim that the majority, and even all, SNIPs come from the CD scenario.   

The new observations that lead to the new expression for the CEEDTD in equation (\ref{eq:NewexpFinal}) seem to suggest that a non-negligible fraction of SNe Ia comes from the CD scenario. There are some recent observational suggestions for the CD scenario (e.g., \citealt{Hsiaoetal2020, Ashalletal2021}). \cite{Luetal2021}, for example, suggest that the under-luminous SN Ia ASASSN-15hy comes from the CD scenario, and \cite{Kasugaetal2021} further mention that Kepler SNR might have come from the CD scenario. 
I find that the new observations of SN 2011fe by \cite{Tuckeretal2021} that reveal the decay of  $^{55}$Fe support the CD scenario for SN 2011fe \citep{Sokeretal2014}, although the DD-MED scenario can also account for SN 2011fe. 
As well, some theoretical studies continue to explore the CD scenario for regular and peculiar SNe Ia (e.g., \citealt{Ablimit2021})

My new claims require further theoretical studies and more observational tests. 
The main theoretical task is to show that some core-WD merger (or WD-WD merger) remnants can live for at least up to $\approx 10^4 \yr$ before they explode. Even longer MED times are possible, but then with decreasing explosion rate. 
The new study by \cite{Neopaneetal2022} is very encouraging. They argue that (1) WD-WD mergers produce a substantial population within a narrow mass range close to the Chandrasekhar mass limit, and (2) the MED time might be as long as $\approx 100 \yr$ and more (see also \citealt{IlkovSoker2012}). The next theoretical stage would be to improve the core-WD numerical simulations of \cite{AznarSiguanetal2015} and to show that this merger process can also lead to a large population of WDs with masses close to the Chandrasekhar mass limit as the CD scenario requires (e.g., \citealt{Ilkov2013}) and try to reach MED times of at least up to $t_{\rm CEED}=t_{\rm MED} \approx 10^4 \yr$. Note that the CD scenario allows also for much longer $t_{\rm MED}$ \citep{IlkovSoker2012},  in addition to the short $t_{\rm MED}$ that is the focus of this study. 

Observations can also support or reject my new suggestion for the CEEDTD. 
The present results suggest, for example, that $\approx 50\%$ of all SNe Ia should experience ejecta-CSM interaction within $t \la 30-100 \yr$ of explosion. This interaction might lead to a re-brightening in different bands, e.g., optical and UV (e.g., \citealt{Grahametal2019}), X-ray, and radio. This calls for a long-time monitoring of SNe Ia. 
As I commented in \cite{Soker2019CEEDTD} the CSM might contain large amounts of dust that by light reflection can cause re-brightening months to years after explosion. \cite{Maund2019} raised the possibility that the few years post-explosion delayed re-brightening of the type IIb SN~2011dh might come from a light echo. In SNe Ia the dust mass is lower than in core collapse supernovae, and so the re-brightening will be faint. Some SNe Ia do indeed show light echo (e.g, \citealt{Graur2019}). 
The CD scenario suggests that in some cases the echoing dust will have elliptical and bipolar morphologies as many PNe do, although in most cases the halos are more spherical (e.g., \citealt{Corradietal2003}).

\section*{Acknowledgments}

 I thank an anonymous referee for useful comments.  This research was supported by a grant from the Israel Science Foundation (769/20).



\end{document}